\documentclass[final,3p,times,twocolumn]{elsarticle}

\usepackage{amssymb}

\pdfoutput=1
\usepackage[table,xcdraw]{xcolor}
\usepackage{graphicx}
\usepackage{epstopdf}
\usepackage{mathrsfs}
\usepackage{amssymb}
\usepackage{verbatim}
\usepackage{color}
\usepackage{hhline}
\usepackage{stackrel}
\usepackage[pdfencoding=auto]{hyperref}
\usepackage{array,multirow}
\usepackage{colortbl}
\usepackage{hhline}
\usepackage{lipsum, babel}
\usepackage{xcolor}
\usepackage{amsmath}
\usepackage{multirow}
\usepackage[table,xcdraw]{xcolor}
\usepackage{slashed}
\usepackage{lipsum}

\usepackage{mathtools, cuted}

\allowdisplaybreaks

\newcommand{\genericT}{\ensuremath{T}}

\newcommand{\mptvec}{\slashed{\vec{p}}_\genericT}

\newcommand{\beq}{\begin{equation}}
\newcommand{\eeq}{\end{equation}}

\journal{Physics Letters B}

\biboptions{sort&compress}

\begin{document}

\begin{frontmatter}



\title{\texorpdfstring{$M_{TN}$}{MTN} is all you need:
production of multiple semi-invisible resonances at hadron colliders}


\author[UK]{Zhongtian Dong\fnref{contribution}} 
\author[UK]{Kyoungchul Kong\fnref{contribution}}
\author[UF]{Konstantin T.~Matchev\fnref{contribution}}
\author[UF]{Katia Matcheva\fnref{contribution}}

\fntext[contribution]{All authors share equal contributions to this paper.}
\affiliation[UK]{organization={Department of Physics and Astronomy, University of Kansas},
            city={Lawrence},
            state={KS},
            postcode={66045}, 
            country={USA}}
            
\affiliation[UF]{organization={Institute for Fundamental Theory, Physics Department, University of Florida},
            city={Gainesville},
            state={FL},
            postcode={32611}, 
            country={USA}}

\begin{abstract}
The stransverse mass variable $M_{T2}$ was originally proposed for the study of hadron collider events in which $N=2$ parent particles are produced and then decay semi-invisibly. Here we consider the generalization to the case of $N\ge 3$ semi-invisibly decaying parent particles. We introduce the corresponding class of kinematic variables $M_{TN}$ and illustrate their mathematical properties. Many of the celebrated features of the $M_{T2}$ kinematic endpoint are retained in this more general case, including the ability to measure the mass of the invisible daughter particle from the stransverse mass kink. We describe and validate a numerical procedure for computing $M_{TN}$ in practice. We also identify the configurations of visible momenta which result in nontrivial ($M_{TN}\ne 0$) values, and derive a pure phase-space estimate for the fraction of such events for any $N$.
\end{abstract}




\end{frontmatter}


\section{Introduction}
\label{sec:intro}

The dark matter problem continues to be among the most compelling arguments for particles and interactions beyond the Standard Model (BSM). In high energy colliders like the Large Hadron Collider (LHC) at CERN, dark matter particles, once produced, will escape the detector, leaving a smoking gun signature of missing transverse momentum $\mptvec$. While theoretically appealing, $\mptvec$ signatures at hadron colliders pose a significant challenge when it comes to measuring the masses of the BSM particles in the event --- the presence of invisible daughter particles prevents the direct reconstruction of semi-invisible parent resonances.

Just over 40 years ago, the transverse mass $M_T$ \cite{Smith:1983aa,Barger:1983wf} was invented to handle the single production of a semi-invisible resonance, in that case a leptonically decaying $W$-boson \cite{UA1:1983crd,UA2:1983tsx}. However, in typical BSM models with dark matter candidates the new particles are expected to be produced in pairs, due to a conserved parity stabilizing the dark matter lifetime. Some well-known examples include: R-parity in low-energy supersymmetry \cite{Jungman:1995df,Feng:2005ee}, Kaluza-Klein parity in models with extra dimensions \cite{Appelquist:2000nn,Cheng:2002ab,Agashe:2007jb}, T-parity in Little Higgs models \cite{Cheng:2003ju,Low:2004xc,Birkedal:2006fz}, U-parity as a discrete remnant of a gauge symmetry \cite{Hur:2007ur,Lee:2008pc}, etc. To handle such cases of pair-production, the Cambridge $M_{T2}$ variable\footnote{Initially the most common application of $M_{T2}$ was in searches for ``sparticles" in supersymmetry, hence $M_{T2}$ also became known as the ``stransverse" mass \cite{Cho:2007qv}.} was introduced close to 25 years ago \cite{Lester:1999tx} and since then has played a major role at the LHC in all kinds of new physics searches \cite{ATLAS:2011hnu,CMS:2012jyl,ATLAS:2013czi,CMS:2015flg,CMS:2016eju,CMS:2017okm,ATLAS:2017vat,ATLAS:2019lff,CMS:2020bfa}, as well as for SM parameter measurements, e.g. in the top sector \cite{CDF:2009zjw,CMS:2013wbt,ATLAS:2012poa,Phan:2013trw,CMS:2017znf} or in the $W$ sector \cite{CMS:2019jcb,ATLAS:2022jat,CMS:2022pio}.

The original $M_{T2}$ proposal \cite{Lester:1999tx} spurred a lot of activity on the theory side as well. For one, $M_{T2}$ was a radically new idea --- up to then, all kinematic variables used in collider physics were known analytical functions of the measured particle momenta and energies. In contrast, $M_{T2}$ was defined algorithmically, and to compute its value, one has to solve an optimization problem involving the unknown invisible momenta. This deceptively simple problem has so far defied all attempts (even employing artificial intelligence in the form of symbolic regression \cite{Dong:2022trn}) to obtain a general  event-by-event analytical formula for $M_{T2}$. While formulas do exist in the literature, they are applicable only for certain special cases \cite{Barr:2003rg,Lester:2007fq,Cho:2007dh,Lester:2011nj,Lally:2012uj}, and in general one has to compute $M_{T2}$ with some sort of numerical minimization software \cite{MT2library,Cheng:2008hk,Walker:2013uxa,Lester:2014yga,Cho:2015laa,Lally:2015xfa,Park:2020bsu}. 

Nevertheless, $M_{T2}$ held great physics potential and was soon applied to many interesting event topologies involving the pair-production of semi-invisibly decaying parent particles. This included situations where $M_{T2}$ is applied to only a subsystem of the event \cite{Kawagoe:2004rz,Burns:2008va}, where the two parents are different (or decay differently) \cite{Barr:2009jv,Konar:2009qr}, where there are multiple invisible daughters per decay chain \cite{Barr:2003rg,Agashe:2010tu,Mahbubani:2012kx}, etc. While $M_{T2}$ is an inherently two-dimensional object defined in the transverse plane, there are useful one-dimensional \cite{Konar:2009wn} and three-dimensional \cite{Cho:2014yma,Cho:2014naa} variants of it as well. A complete review of the vast literature on $M_{T2}$ is beyond the scope of this letter and the interested reader is referred to any one of the many nice reviews \cite{Barr:2010zj,Barr:2011xt,Lester:2013aaa,Matchev:2019sqa,Franceschini:2022vck}.

In this letter we consider the generalization of $M_{T2}$ to the corresponding $M_{TN}$ kinematic variable applicable in the case of producing multiple ($N\ge 3$) parent particles ${\cal P}_1$, ${\cal P}_2$, $\ldots$, ${\cal P}_N$, decaying semi-invisibly as in Fig.~\ref{fig:event_topology}.\footnote{Our $M_{TN}$ should not be confused with the $M_{TX}$ variable defined in \cite{Barr:2003rg}, where $X$ indicated the number of invisible daughter particles resulting from the production of {\em two} parents ${\cal P}_1$ and ${\cal P}_2$.}  Our interest in this scenario is driven by several reasons:
\begin{itemize}
\item {\em Aesthetics.} We shall encounter some beautiful math which was not present in the case of $N=2$.
\item {\em Open-mindedness.} The assumption that the parents are pair-produced is certainly not ironclad and can be relaxed with some clever model-building. 
\item {\em Pragmatism.} Apart from BSM searches, the LHC will eventually be sensitive to rare SM processes of multi-top and/or multi-$W$ production in purely leptonic channels. Such event topologies fit Fig.~\ref{fig:event_topology} and can be probed with the new class of kinematic variables. For example, CMS and ATLAS have already observed triple $W$ production \cite{CMS:2020hjs,ATLAS:2022xnu} and four-top-quark production \cite{ATLAS:2023ajo,CMS:2023ftu}, the next targets being three top-quarks \cite{Barger:2010uw,Cao:2019qrb,Khanpour:2019qnw,Boos:2021yat}
and four $W$ bosons \cite{Barger:1989cp}.
\end{itemize}

The paper is organized as follows. In Section~\ref{sec:mtn_definition} we introduce the $M_{TN}$ variable and in Section~\ref{sec:mtn_properties} we discuss some of its mathematical properties. In Section~\ref{sec:numerical} we describe and validate our numerical procedure for computing $M_{TN}$. In Section~\ref{sec:mg5} we illustrate typical applications of $M_{TN}$ to realistic physics examples, namely $WWW$ and $WWWW$ production at the LHC in purely leptonic channels, where the relevant kinematic variables are $M_{T3}$ and $M_{T4}$, respectively. In Section~\ref{sec:kink} we prove the existence of a kink structure at the correct value of the daughter mass by deriving the corresponding analytical expressions for the parent-daughter mass relationship following from the experimental measurement of the upper kinematic endpoint in the $M_{TN}$ distribution. Section~\ref{sec:conclusion} is reserved for our conclusions.

\section{Definition of \texorpdfstring{$M_{TN}$}{MTN}}
\label{sec:mtn_definition}

\begin{figure}[t!]
    \centering
    \includegraphics[width=0.4\textwidth]{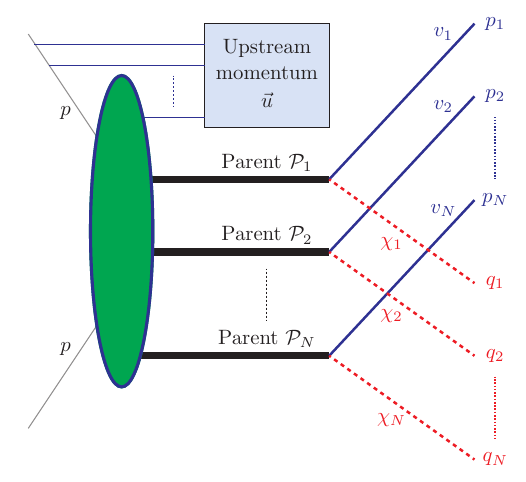} 
    \caption{The event topology under consideration in this paper. The blue solid lines denote SM particles visible in the detector, e.g., leptons, photons or jets. The red dashed lines correspond to final state particles invisible in the detector (neutrinos or dark matter candidates). The blue box indicates visible particles radiated either from the initial state or from decays upstream.}
    \label{fig:event_topology}
\end{figure}

As depicted in Fig.~\ref{fig:event_topology}, we consider the inclusive production of $N$ identical parent particles, ${\cal P}_i$ $(i=1,2,\ldots,N)$, at a proton-proton ($pp$) collider like the LHC. Each parent particle ${\cal P}_i$ decays to a visible particle  $v_i$ with mass $m_i$ and 4-momentum $p_i = (E_i, \vec p_{i T}, p_{i z})$ and an invisible particle $\chi_i$ with 4-momentum $q_i=(\varepsilon_i, \vec q_{i T}, q_{i z})$. For simplicity, we shall assume that the invisible particles $\chi_i$ are identical and have a common mass $m_\chi$. Since the actual mass $m_\chi$ is apriori unknown, in all calculations below we shall need to use an ansatz for it which will be denoted with a tilde: $\tilde m_\chi$. The individual invisible particle momenta $\vec{q}_i$ are not measured in the detector, but momentum conservation in the transverse plane implies
\begin{equation}
\sum_{i=1}^N \vec{q}_{iT} = \mptvec \equiv - \sum_{i=1}^N \vec{p}_{iT} - \vec{u}_{T}\, ,
\label{eq:mptdef}
\end{equation}
where we allow for the presence of additional visible radiation in the event with total transverse momentum $\vec{u}_T$. This could arise from initial state radiation, or from the decays of even heavier, ``grandparent'', particles. 

The main ingredients in the $M_{TN}$ calculation will be the transverse masses $M_{T{\cal P}_i}$ of the $N$ parent particles ${\cal P}_i$:
\beq
M_{T {\cal P}_i}(\vec q_{iT},\tilde m_\chi) = \sqrt{m_i^2 + \tilde m_\chi^2  
 + 2 \big ( E_{iT} \varepsilon_{iT}  - \vec p_{iT} \cdot \vec q_{iT} \big )},~~~~
\label{eq:MTPdef} 
\eeq
where the transverse energies are defined as
\beq
E_{i T} = \sqrt{ \vec p_{i T}^{\, 2} + m_i^2  } , \quad
\varepsilon_{i T} = \sqrt{ \vec q_{i T}^{\,2} + \tilde m_\chi^2 } \, .
\eeq
In analogy to $M_{T2}$, the $M_{TN}$ variable can be defined as
\begin{equation}
M_{TN} (\tilde m_\chi) \equiv \min_{\substack{\vec{q}_{iT}\\
\sum_i\vec{q}_{iT} = \mptvec }}
\left\{\max_i\left[M_{T{\cal P}_i}(\vec{q}_{iT},\tilde m_\chi) \right] \right\} , \label{eq:MTNdef}
\end{equation}
where the minimization over the individual invisible transverse momenta $\vec{q}_{iT}$ is done subject to the constraint (\ref{eq:mptdef}). 

\section{Mathematical properties of \texorpdfstring{$M_{TN}$}{MTN}}
\label{sec:mtn_properties}

The construction outlined in the previous section guarantees that on an event-by-event basis the computed value of $M_{TN}$ does not exceed the common mass $M_{\cal P}$ of the parents ${\cal P}_i$ (of course, provided that the correct daughter mass $m_\chi$ has been used for the ansatz $\tilde m_\chi$). In other words,
\begin{equation}
M_{TN} (\tilde m_\chi = m_\chi) \le M_{\cal P}.
\label{eq:MTNlimit}
\end{equation}
It is easy to see why this bound is always true --- in the process of minimizing over $\vec{q}_{iT}$ in eq.~(\ref{eq:MTNdef}), we will eventually hit on the {\em actual} values of $\vec{q}_{iT}$ in that event, for which $M_{TN}(m_\chi)=M_{\cal P}$ by construction. From there, exploring invisible momenta away from the actual ones can only further reduce the value of $M_{TN}$. 

In light of the inequality (\ref{eq:MTNlimit}), $M_{TN}$ is a {\em mass-bound} kinematic variable and provides an event-by-event lower bound on the true invariant mass of the parent particle \cite{Cheng:2008hk} (see also the general discussion in section IX.B of \cite{Barr:2011xt}). However, the usefulness of such variables in practice crucially depends on whether the upper bound in (\ref{eq:MTNlimit}) is saturated or not. If the bound is saturated, then the measurement of the upper kinematic endpoint $M_{TN}^{max}$ of the $M_{TN}$ distribution gives the parent mass $M_{\cal P}$ as a function of the ansatz $\tilde m_\chi$: 
\begin{equation}
M_{\cal P}(\tilde m_\chi) = M_{TN}^{max}(\tilde m_\chi).
\label{eq:MTNmax_mchi}
\end{equation}

Fortunately, in our setup here, the bound in (\ref{eq:MTNlimit}) is indeed saturated, and eq.~(\ref{eq:MTNmax_mchi}) holds. This is most easily proven by example, i.e., by finding events with visible momentum configurations for which the minimization over $\vec{q}_{iT}$ picks up the {\em actual} invisible momenta in the event. The one-dimensional colinear momentum configurations discussed below in Section~\ref{sec:kink} are precisely such events --- one can easily check that either one of the two branches in eq.~(\ref{end110PT}) satisfies eq.~(\ref{eq:MTNlimit}) (for a related discussion in the case of $M_{T2}$, see \cite{Matchev:2009fh}).

In what follows, we shall focus on the limit of massless visible particles in the final state, i.e. $m_i = 0, \forall i=1,2,\ldots, N$. This approximation is well motivated for the standard reconstructed objects like electrons, muons, photons and jets, especially at high $p_{iT}$, as required by the triggers. Furthermore, as we have already learned from numerous $M_{T2}$ studies, in the massless limit the math simplifies a lot, while keeping the essential physics unchanged.

\begin{figure}[t!]
    \centering
    \includegraphics[width=0.4\textwidth]{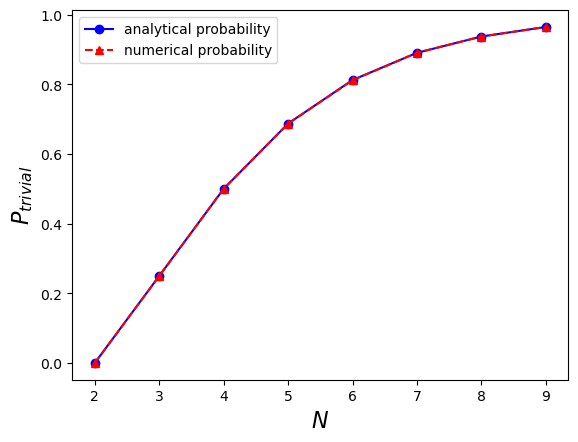}
    \caption{Fraction of events with $M_{TN}=0$ for different values of $N$, in an event sample produced by sequentially sampling the visible transverse momenta according to (\ref{eq:sampling}). The blue circles correspond to our theoretical prediction (\ref{eq:prediction}), while the red triangles represent the numerical result using 1,000,000 events for each value of $N$. 
\label{fig:compare} }
\end{figure}

For $N\ge 3$, the $M_{TN}$ kinematic distributions develop a new feature which is not present in the case of $M_{T2}$: a sizable fraction of events have trivial $M_{TN}$ values ($M_{TN}=0$). We determined that {\em nontrivial} $M_{TN}$ values arise only for visible momentum configurations where all $\vec{p}_{iT}$ vectors are in the same half-plane (see Section~\ref{sec:numerical} for further details). This condition is reminiscent of a problem from elementary geometry (see acknowledgements) which asks how often $N$ points sampled on a circle lie on one semi-circle. If the points on the circle are chosen at random, the answer is given by $2N/N^2$ and therefore the probability $P_{trivial}$ that we get a {\em trivial} solution for $M_{TN}$ is 
\begin{equation}
P_{trivial}(N) = 1- \frac{2N}{2^N}.
\label{eq:prediction}
\end{equation}
This prediction is plotted in Fig.~\ref{fig:compare} with blue circles. Note that for $N=2$, $P_{trivial}=0$ and hence there are no trivial solutions for $M_{T2}$, in agreement with previous observations.

In order to check the validity of the result (\ref{eq:prediction}) numerically, we exhaustively scan the full phase space of the $N$ visible transverse momenta as follows. First, we order the $N$ visible particles by the magnitude of their $\vec{p}_T$:
\begin{equation}
p_{(n)T} \equiv \text{max}_n \left\{ p_{iT}\right\}, \quad n=1,2,\ldots,N\, ,
\end{equation}
where we introduce the notation $\text{max}_n\{\cal S\}$ to denote the $n$-th largest element in a set ${\cal S}$ \cite{Kim:2015bnd}. Without loss of generality, we can set the hardest momentum $p_{(1)T}=1$ and fix its orientation (e.g., along the $x$-axis). Then we sample the transverse momentum components of the remaining visible particles sequentially from uniform distributions $U(a,b)$ as follows:
\begin{equation}
p_{(n)T} \sim U(0,p_{(n-1)T}), ~~ \varphi_{(n)} \sim U(0,2\pi),  ~~n\ge 2.
\label{eq:sampling}
\end{equation}
Since the azimuthal angles $\varphi_{(n)}$ are sampled uniformly, the directions of the transverse momenta are chosen at random and the assumption behind eq.~(\ref{eq:prediction}) holds. Then for each integer value of $N$, we generate 1,000,000 events according to (\ref{eq:sampling}), numerically compute $M_{TN}$ (as described in the next section) and calculate the fraction of events with trivial $M_{TN}$ solutions. The result is shown in Fig.~\ref{fig:compare} with red triangles and is in perfect agreement with the prediction (\ref{eq:prediction}) shown in blue circles.

The exhaustive scan of the visible phase space also allowed us to check numerically whether the solutions for $M_{TN}$ are balanced or not. (A balanced solution is obtained whenever the minimization in (\ref{eq:MTNdef}) results in invisible momenta such that {\em all} transverse parent masses are equal: $M_{T{\cal P}_1}=M_{T{\cal P}_2}=\ldots =M_{T{\cal P}_N}$.) We verified that $M_{TN}$ is {\em always} given by a balanced solution. We note that our conclusions regarding $P_{trivial}$ and the prevalence of the balanced solution persist even when we artificially add upstream transverse momentum $\vec{u}_T$ to the event.

\section{Computation of \texorpdfstring{$M_{TN}$}{MTN}}
\label{sec:numerical}

In this section we describe and validate our numerical code for computing $M_{TN}$. We used the {\sc scipy} module {\sc optimize}, which is designed for minimizing objective functions like our eq.~(\ref{eq:MTNdef}). One subtlety is that in our numerical experiments we noticed that local optimizer options like \verb^Nelder-Mead^ and \verb^COBYLA^ give suboptimal results, so we recommend using a global optimizer option (we chose \verb^differential_evolution^).

\begin{figure}[t!]
    \centering
    \includegraphics[width=0.4\textwidth]{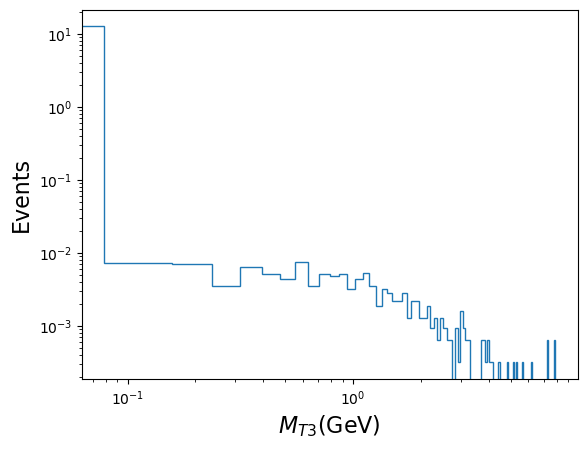} 
    \caption{A unit-normalized histogram of the values of $M_{T3}(\tilde m_\chi=0)$ computed by the {\sc scipy.optimize} module for purely leptonic $WWW$ events in which the momentum configuration corresponds to  a trivial solution ($M_{T3}=0$). The search range for the invisible momentum components was set to $\pm 5$ TeV.
    }
    \label{fig:compare_MT3equal0}
\end{figure}

In order to assess the accuracy of the {\sc scipy.optimize} minimization, we performed two tests. In each case, we compare the output from the optimizer to the theoretical expectation. Unfortunately, in the absence of a general analytical formula for $M_{TN}$, we are limited in what kinds of tests we can do. One possibility, illustrated in Fig.~\ref{fig:compare_MT3equal0}, is to restrict ourselves only to momentum configurations which we know (see discussion in Section~\ref{sec:mtn_properties}) should give trivial solutions ($M_{TN}=0$) and then check how close to zero the result from the optimizer can get. For this exercise we use purely leptonic $WWW$ events generated with {\sc MadGraph5\textunderscore}a{\sc MC@NLO} \cite{Alwall:2011uj} and show a histogram of the computed $M_{T3}$ values in Fig.~\ref{fig:compare_MT3equal0}. As expected, the large majority of events indeed fall in the zero bin of trivial $M_{T3}$ values. However, we also observe a tail of non-zero $M_{T3}$ values which can be as large as a few GeV, hinting at an apparent problem with the minimization. We have tracked the origin of this tail to the finite search range of invisible momenta --- in creating Fig.~\ref{fig:compare_MT3equal0} the range was set to $\pm 5$ TeV. However, the minimum of the objective function (\ref{eq:MTNdef}) could easily end up outside this range. To see this, consider the following visible momentum configuration (in Cartesian coordinates): $\vec{p}_{1T}=(a,0)$, $\vec{p}_{2T}=(0,b)$, $\vec{p}_{3T}=(-c,-\varepsilon)$, where $a$, $b$ and $c$ are some finite positive constants, while $\varepsilon$ is a positive infinitesimal parameter ($a\sim b \sim c \gg \varepsilon > 0$). As discussed in the previous section, this is an example of a trivial ($M_{T3}=0$) event since the momenta are not in the same half-plane. It is easy to verify that the smallest (in magnitude) invisible momenta minimizing the objective function are as follows: $\vec{q}_{1T}=(\frac{bc}{\varepsilon}-a,0)=(\frac{bc}{\varepsilon a}-1)\, \vec{p}_{1T}$, $\vec{q}_{2T}=(0,0)$ and $\vec{q}_{3T}=(-\frac{bc}{\varepsilon}+c,-b+\varepsilon)=(\frac{b}{\varepsilon}-1)\, \vec{p}_{3T}$. Since $\vec{q}_{1T}$ is colinear with $\vec{p}_{1T}$, $\vec{q}_{3T}$ is colinear with $\vec{p}_{3T}$, and $\vec{q}_{2T}=0$, all three transverse parent masses $M_{T{\cal P}_i}$ vanish and $M_{T3}=0$, as expected. However, note that as $\varepsilon \to 0$, the magnitudes of $\vec{q}_{1T}$ and $\vec{q}_{3T}$ grow indefinitely and will eventually fall outside the search region, no matter how large we set it from the beginning. Therefore, any numerical minimization routine will have a difficult time finding such solutions, especially as $\varepsilon$ approaches zero. The brute force solution to this problem is to keep enlarging the search region whenever the solution for some invisible momentum ends up on the boundary (we have checked that upon enlarging the search region, the tail seen in Fig.~\ref{fig:compare_MT3equal0} gradually shrinks). However, the discussion from the previous section already suggests a much more elegant and robust solution --- before attempting any numerical minimization, one should first check whether the momentum configuration leads to a non-trivial solution, and only in that case proceed with the minimization; otherwise, return $M_{TN}=0$ and bypass the numerical minimization altogether.

\begin{figure}[t!]
    \centering
    \includegraphics[width=0.4\textwidth]{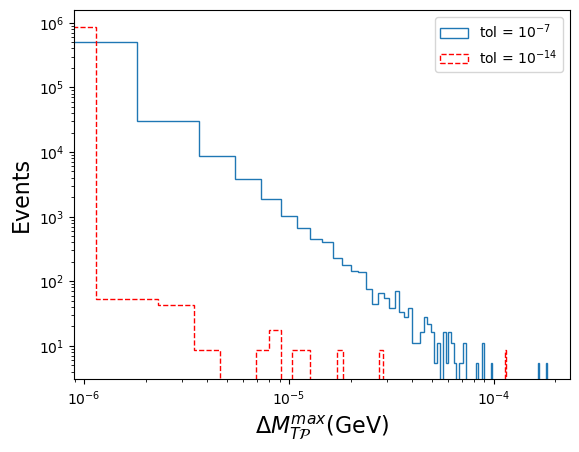}
    \caption{Unit-normalized distribution of the maximal absolute difference (\ref{eq:deltaMTPmax}) between the transverse masses of two parent particles ${\cal P}_i$ and ${\cal P}_j$ in purely leptonic $WWW$ events with nontrivial $M_{T3}$ values. The test mass was taken to be $\tilde m_\chi=0$. Results are shown for two different values of the numerical tolerance: $10^{-7}$ (blue solid histogram) and $10^{-14}$ (red dashed histogram).
    }
    \label{fig:compare_deltaMT}
\end{figure}

In the second test, we make use of the property discussed in the previous section, namely, that $M_{TN}$ always results in a balanced solution. Therefore, this time we choose leptonic $WWW$ events whose visible momentum configurations correspond to nontrivial $M_{T3}$ values, and then check the maximal transverse mass difference between any pair of parents: 
\begin{equation}
\Delta M_{T{\cal P}} ^{max}\equiv \max_{i,j=1,2,3}|M_{T{\cal P}_i}-M_{T{\cal P}_j}|. 
\label{eq:deltaMTPmax}
\end{equation}
The result is histogrammed in Fig.~\ref{fig:compare_deltaMT} for two different values of the tolerance: $10^{-7}$ (blue histogram) and $10^{-14}$ (red histogram). For definiteness, we chose the invisible mass ansatz to be $\tilde m_\chi=0$. We observe that the precision on the $M_{T3}$ calculation is pretty good --- below $\sim$ 0.1 MeV --- and can be further improved, if necessary, by tightening the tolerance.

\section{LHC examples of \texorpdfstring{$M_{TN}$}{MTN} usage} 
\label{sec:mg5}

In order to showcase typical applications of the new $M_{TN}$ variables in collider physics, we consider purely leptonic channels of multi-$W$ production at the LHC. Events were generated with {\sc MadGraph5\textunderscore}a{\sc MC@NLO} \cite{Alwall:2011uj} at a $pp$ collider of 14 TeV center-of-mass beam energy. We consider three processes --- $W$ pair-production, for which the relevant variable is $M_{T2}$, triple $W$ boson production, where $M_{T3}$ is appropriate, and finally, quadruple $W$ boson production, which can be probled with $M_{T4}$. For simplicity, we consider exclusive production, i.e., all events have $\vec{u}_T=0$. Each $W$ boson is forced to decay to a lepton (electron or muon), thus the final state signature is $N$ leptons plus $\mptvec$.

\begin{figure*}[t!]
    \centering
    \includegraphics[width=0.33\textwidth]{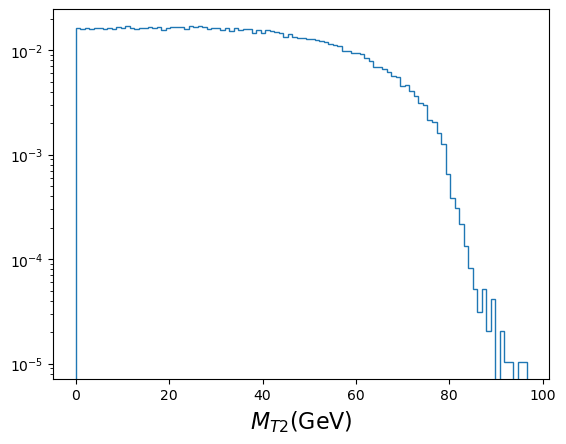}
    \hspace*{-0.2cm}
    \includegraphics[width=0.33\textwidth]{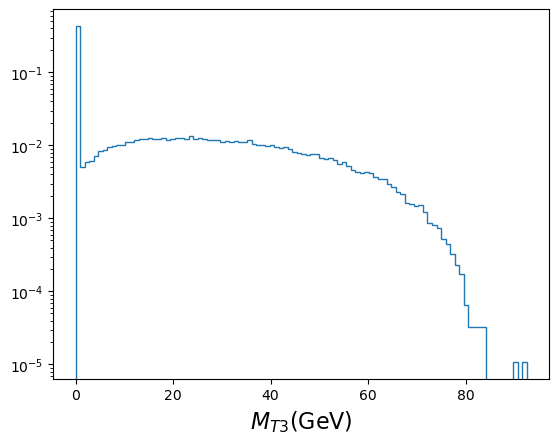} 
    \hspace*{-0.2cm}
    \includegraphics[width=0.33\textwidth]{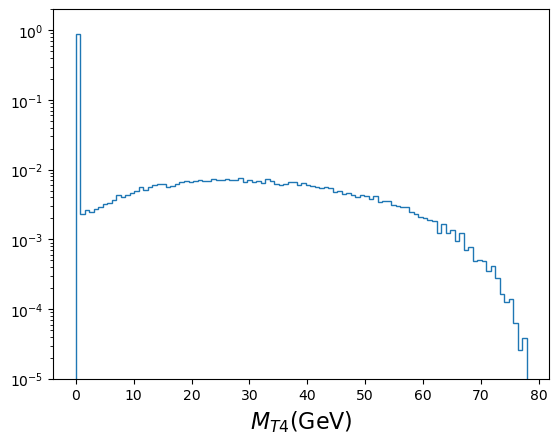}
    \caption{Parton-level unit-normalized distributions of: $M_{T2}$ in leptonic $WW$ events (left panel), $M_{T3}$ in leptonic $WWW$ events (middle panel) and $M_{T4}$ in leptonic $WWWW$ events at the LHC. }
    \label{fig:Wevents}
\end{figure*}

We then proceed to calculate the relevant $M_{TN}$ variable as explained in the previous section. The resulting unit-normalized distributions for $M_{T2}$, $M_{T3}$ and $M_{T4}$ are shown in the left, middle and right panel of Fig.~\ref{fig:Wevents}, respectively. In each case, the test mass was chosen to be the true mass of the daughter neutrino, i.e., $\tilde m_\chi=0$.

Comparing the three panels in Fig.~\ref{fig:Wevents}, we immediately notice the spike at $M_{T3}=0$ and $M_{T4}=0$, which is absent in the $M_{T2}$ distribution. As already discussed in Section~\ref{sec:mtn_properties}, this is a new feature for $N\ge 3$, and is due to the presence of a sizable number of visible momentum configurations which result in zero $M_{TN}$ values (see Figure~\ref{fig:compare}). Nevertheless, in all three cases, the $M_{TN}$ distributions do their job of determining the parent mass from the upper kinematic endpoint according to (\ref{eq:MTNmax_mchi}). The spillover of events beyond the nominal $W$ mass is due to finite width effects. This suggests that in order to have a reliable measurement of the $M_{TN}$ upper kinematic endpoint in practice, one has to account for both the resonance width effect and the detector resolution.

As a final comment, note that the spike at zero is slightly taller in the case of $M_{T4}$ than $M_{T3}$ --- we find that in the case of $W$ production, $P_{trivial}=39.7\%$ for $N=3$ and $P_{trivial}=67.6\%$ for $N=4$. Due to the effects of the matrix element and the parton distribution functions, $P_{trivial}$ is larger than our simplified prediction (\ref{eq:prediction}) which was derived based on pure phase space. In realistic events, the directions of the hardest and second-hardest lepton are anti-correlated, which increases the chances of a trivial momentum configuration.
A taller spike at zero in turn implies relatively fewer events near the upper kinematic endpoint, causing the endpoint measurement to be statistically limited at higher $N$.

\section{Kink structure}
\label{sec:kink}

A celebrated feature of $M_{T2}$ is the existence of a kink (i.e., a discontinuity in the gradient) in the relationship (\ref{eq:MTNmax_mchi}) derived from the $M_{T2}$ kinematic endpoint \cite{Cho:2007qv,Gripaios:2007is,Barr:2007hy,Cho:2007dh}. The location of the kink is precisely at the true value of the daughter mass, $\tilde m_\chi=m_\chi$, which in principle allows for the determination of the masses of {\em all} the particles involved in the process.

The mathematics behind the $M_{T2}$ kink can be easily extended to the case of $M_{TN}$, as we shall now show. In the massless case $m_i=0$ considered here, the kink arises due to non-vanishing upstream transverse momentum $\vec{u}_T \ne 0$ \cite{Burns:2008va}. For any given choice of the test mass $\tilde m_\chi$, there are two extreme kinematic configurations of the visible momenta which are potentially responsible for the  $M_{TN}$ kinematic endpoint \cite{Burns:2008va,Matchev:2009fh}. Both configurations are purely one-dimensional, i.e., all relevant transverse vectors $\vec{p}_{1T}, \vec{p}_{2T}, \cdots, \vec{p}_{NT}$ and $\vec{u}_T$ are aligned along a common axis in the transverse plane. In the first configuration, responsible for the ``left" side of the relation (\ref{eq:MTNmax_mchi}), the visible momenta $\vec{p}_{iT}$ are all equal to each other, are parallel to $\vec{u}_T$, and have magnitude
\begin{equation}
p_{TL} \equiv 
\frac{M_{\cal P}^2-m_\chi^2}{2M_{\cal P}}
\left( \sqrt{ 1 +\left(\frac{u_T}{NM_{\cal P}}\right)^2} - \frac{u_T}{NM_{\cal P}} \right) \, .
\label{eq:ptLdef}
\end{equation}
In the second configuration, giving the ``right" side of (\ref{eq:MTNmax_mchi}), the visible momenta $\vec{p}_{iT}$ are all equal to each other, anti-parallel to $\vec{u}_T$, and have magnitude
\begin{equation}
p_{TR} \equiv 
\frac{M_{\cal P}^2-m_\chi^2}{2M_{\cal P}}
\left( \sqrt{ 1 +\left(\frac{u_T}{NM_{\cal P}}\right)^2} + \frac{u_T}{NM_{\cal P}} \right) \, .
\label{eq:ptRdef}
\end{equation}

As a result, the are two branches in the function (\ref{eq:MTNmax_mchi}) relating the measured endpoint $M_{TN}^{max}$ to the assumed test mass \cite{Burns:2008va}:
\begin{equation}
M_{TN}^{max}(\tilde m_\chi) =\left\{   
\begin{array}{ll}
F_{L}(\tilde m_\chi)\, , & ~~~{\rm if}\ \tilde m_\chi \le m_\chi\, , \\ [2mm]
F_{R}(\tilde m_\chi)\, , & ~~~{\rm if}\ \tilde m_\chi \ge m_\chi\, , 
\end{array}
\right.
\label{end110PT} 
\end{equation}
where 
\begin{align}
F_{L}(\tilde m_\chi) &=
\left\{ 
\left[
p_{TL} + \sqrt{ \left(p_{TL}+\frac{u_T}{N}\right)^2 + \tilde m_\chi^2} 
\, \right]^2
- \left(\frac{u_T}{N}\right)^2   
\right\}^{\frac{1}{2}}, 
\label{FL110} 
\\
F_{R}(\tilde m_\chi) &=
\left\{ \left[
p_{TR} + \sqrt{ \left(p_{TR}-\frac{u_T}{N}\right)^2 + \tilde m_\chi^2} 
\, \right]^2
- \left(\frac{u_T}{N}\right)^2   \right\}^{\frac{1}{2}},
\label{FR110}
\end{align}
Note that the two branches are simply related as
\begin{equation}
F_{R}(\tilde m_\chi,u_T) = F_{L}(\tilde m_\chi,-u_T),
\end{equation}
which can be easily verified by inspecting eqs.~(\ref{eq:ptLdef}), (\ref{eq:ptRdef}), (\ref{FL110}) and (\ref{FR110}).

\begin{figure}[t!]
    \centering
    \includegraphics[width=0.45\textwidth]{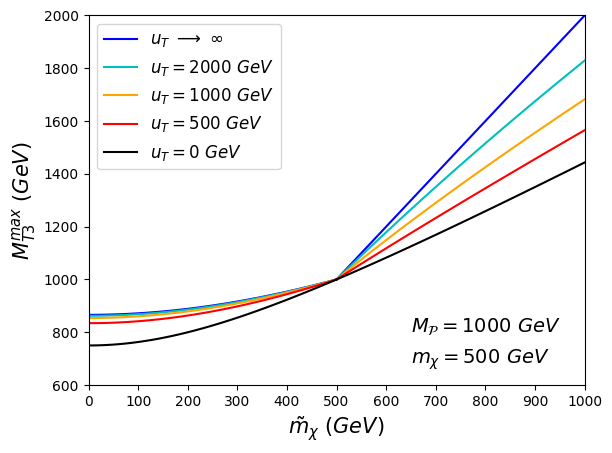}
    \caption{The upper kinematic endpoint $M_{T3}^{max}$ of the $M_{T3}$ distribution for a case with parent mass $M_{\cal P}=1$ TeV and daughter mass $m_\chi=500$ GeV. The lines are color-coded by the values of the upstream transverse momentum $u_T$: no upstream momentum (black), 
    $u_T=500$ GeV (red),
    $u_T=1000$ GeV (orange), and
    $u_T=2000$ GeV (cyan). The blue line represents the limiting case of $u_T\to\infty$.}
    \label{fig:MT3max}
\end{figure}

The function (\ref{end110PT}) is illustrated in Fig.~\ref{fig:MT3max}, for the case of $N=3$, with true parent mass $M_{\cal P}=1$ TeV and daughter mass $m_\chi=500$ GeV. The plot shows the dependence on the test mass $\tilde m_\chi$ for several different values of the upstream transverse momentum: 
$u_T=0$ (black line),
$u_T=500$ GeV (red line),
$u_T=1$ TeV (orange line), and
$u_T=2$ TeV (cyan line). The blue solid line shows the limiting case of $u_T\to\infty$.

The most striking feature of Fig.~\ref{fig:MT3max} is the presence of the kink at precisely the correct daughter mass ($\tilde m_\chi=m_\chi$). 
The sharpness of the kink depends on the size of the upstream transverse momentum --- the larger the $u_T$, the more pronounced the kink.

\section{Conclusions}
\label{sec:conclusion}

With its continuing operations planned for the next 20 years, the LHC is entering an era when it will be sensitive to the simultaneous production of $N>2$ heavy resonances. In many BSM scenarios motivated by the dark matter problem, these resonances are expected to decay semi-invisibly, giving rise to signatures with missing transverse momentum. Additional impetus for studying such event topologies is provided by the leptonic decays of the top quark and $W$ boson in the Standard Model. In this letter we generalized the $M_{T2}$ concept to the case of $M_{TN}$ kinematic variables which are ideally suited for the analysis of these scenarios. We showed that the measurement of the upper kinematic endpoint in the $M_{TN}$ distribution provides one constraint on the masses of the parent and daughter particles, while the kink structure (see Fig.~\ref{fig:MT3max}) is able to pin down those masses completely. We also identified the visible momentum configurations which correspond to $M_{TN}=0$, thus avoiding the need to numerically compute $M_{TN}$ in those cases.

The $M_{TN}$ idea is ripe for further investigations, following the same roadmap as in the $M_{T2}$ case. For example, one can easily generalize the definition (\ref{eq:MTNdef}) to 3+1 dimensions along the lines of \cite{Cho:2014naa,Cho:2014yma,Cho:2015laa}. It is also worth exploring the properties of $M_{TN}$ beyond the massless approximation and for asymmetric event topologies, e.g., with different mother \cite{Barr:2009jv} or daughter \cite{Konar:2009qr} particles. The $M_{TN}$ variables can also be used as input features in deep learning multi-variate approaches to LHC data analysis. One could even use deep learning to test whether $M_{TN}$ are indeed optimal variables for their respective event topologies \cite{Kim:2021pcz}. In any case, we are hoping that some $M_{TN}$ will soon play a role in a new physics discovery at the LHC.

\section*{Acknowledgments}
We thank the Buchholz High School math team coach Ziwei~Lu and math team member Philip Matchev for pointing out the connection to the points-on-a-circle math problem and sharing valuable insights about its solution. 
This work is supported in part by the U.S. Department of Energy award number DE-SC0022148, DE-SC0024407 and DE-SC0024673. 
ZD is supported in part by College of Liberal Arts and Sciences Research Fund at the University of Kansas.

\bibliographystyle{elsarticle-num}
\bibliography{draft}

\end{document}